\documentclass{ws-ijmpa}

\begin{document}

\markboth{GHISELLINI}{EXTRAGALACTIC GAMMA-RAYS}

\catchline{}{}{}{}{}

\title{EXTRAGALACTIC GAMMA--RAYS: GAMMA RAY BURSTS AND BLAZARS}

\author{GABRIELE GHISELLINI}
\address{Osservatorio Astronomico di Brera. Via Bianchi 46 Merate, I--23807 Italy;
gabriele@merate.mi.astro.it}

\maketitle


\begin{abstract}
The extragalactic gamma--ray sky is dominated by two classes of
sources: Gamma--Ray Bursts (GRBs) and radio loud active galactic 
nuclei whose jets are pointing at us (blazars).
We believe that the radiation we receive from them originates from
the transformation of bulk relativistic energy into random energy.
Although the mechanisms to produce, collimate and accelerate the jets
in these sources are uncertain, it is fruitful to compare
the characteristics of both classes of sources in search
of enlightening similarities.
I will review some general characteristics of radio loud AGNs
and GRBs and I will discuss the possibility that both classes of sources
can work in the same way.
Finally, I will discuss some recent exciting prospects to use
blazars to put contraints on the cosmic IR-Optical-UV backgrounds,
and to use GRBs as standard candles to measure the Universe.
\end{abstract}



\section{Introduction}

The golden `60s, together with the Cosmic Microwave Background and
the pulsars, witnessed the discovery of other two
important classes of sources: Gamma--Ray Bursts and blazars
(i.e. BL Lacertae objects and Flat Spectrum Radio Quasars).
GRB are flashes of $\gamma$--ray radiation, likely flagging the
birth of a stellar size black hole, while radio--loud AGNs, even if 
remarkable for their rapid variability, live for hundreds millions
years, producing spectacular and Mpc--size jets and radio lobes,
and are powered by supermassive ($10^6$--$10^{10}M_\odot$) black holes.
This apparent diversity should not outshine some remarkable similarities,
since in both classes the emitting plasma is moving
at relativistic bulk speeds, and the radiation we see is likely
the result of the transformation of part of this well ordered
kinetic energy into random energy and then into radiation.
Furthermore, there are strong evidences that also GBRs have 
collimated jets.
Very different at first sight, GRBs and blazars might be powered
by a similar machine.
Consider also that the dynamical timescale for a GRB
should be of the order of the light travel time to cross the gravitational
radius, i.e. $R_{\rm g}/c\sim 10^{-5}M/M_\odot$ seconds.
A burst with a duration of 10 seconds therefore lasts for $10^6$ 
dynamical times: it can be a quasi steady process (for a $10^9$ solar mass
black hole, this time is equivalenth to 300 years).
What we naively consider an ``explosion" is instead a long event.

In both classes of sources we see non--thermal radiation
and it is therefore likely  
that magnetic field and non--thermal particles
are the key ingredients to produce the radiation we see.
This radiation, being produced by plasma in relativistic motion,
is strongly beamed in the velocity direction, and we have evidences
that also in GRBs the emitting fireball is collimated in a cone,
i.e. a ``jet".
For these reasons it is instructive to compare them looking for
similarities and differences, to see if their physics is similar.
In the following I will briefly discuss some of the basic facts
of blazars and GRBs, and discuss the possibility that, at the origin
of their phenomenology, there is a common engine.
Being so powerful, blazars and GRBs are well visible even at redshifts
greater than 6, and we can use them as probes of the far universe.
In the case of GRBs, we can even {\it measure} the universe, 
thanks to a recently found correlation between their properties
which enable us to use them as standard candles.

\section{Blazars vs GRBs}

\subsection{Speeds: blazars}

Soon after the achievement of the VLBI technology, in the early `70s, 
3C 279 and 3C 273 were discovered to have spots of radio emission
moving away from the nucleous at the apparent speed of 
$\beta_{\rm app}\sim$10.
This was one of the few dramatic features in astronomy which were
anticipated by theory (Rees 1966), and it is the result of real motion
at velocities close to $c$, observed at small, albeit not vanishingly
small, viewing angles.
The corresponding bulk Lorentz factors $\Gamma$ are greater or equal than
$\beta_{\rm app}$, and today superluminal speeds are routinely measured
in almost all (and the exception are important) blazars.
It is diffucult to assign an upper limit to the inferred distribution 
of $\Gamma$, but it is believed that in powerful blazars
(showing relatively strong broad emission lines, as in radio--quiet AGNs),
the range is $3<\Gamma<20$, with some remaining uncertainty 
associated with the phenomenon of intraday variability in radio, 
leading to huge brightness temperatures
(see Wagner \& Witzel 1995 for a review).
This is now explained by the effect interstellar scintillation, but
a source scintillates if its angular size is small, suggesting in any
case very high brightness temperatures, even if not so extreme,
requiring large $\Gamma$ in any case (e.g. Kedziora--Chudczer et al. 2001).
Furthermore, at the other hand of the power scale, models of TeV BL Lacs 
seem to require $\Gamma$ up to $\sim$50 or so, to explain 
their TeV emission, especially when their $\gamma$--ray spectrum is 
``de--absorbed", i.e. when the depletion of TeV photons interacting
with the IR cosmic background is taken into account 
(e.g. Krawczynski, Coppi \& Aharonian 2002; 
Konopelko et al. 2003).
Interestingly, VLBI and space VLBI observations of these TeV BL Lacs 
have revealed that their apparent speed at the submas scale ($\sim$1 pc)
is only moderately superluminal or even subluminal (Piner \&
Edwards 2004; Giroletti et al. 2004).
This of course opens the issue of strong deceleration from TeV emitting
region ($\sim 10^{17}$ cm from the center) to 1 pc 
(Kazanas \& Georganopoulos 2003, 
Ghisellini, Tavecchio \& Chiaberge 2004).

From the beginning, the plasma moving inside the jets was assumed
to have a monodirectional velocity, despite the conical geometry
of the jet itself.
There have been sporadic attempts to see what happens if the velocity
vectors are distributed in a fan--like geometry (``sprayed", 
Celotti et al. 1993), but this possibility has not been followed in any detail.
Instead, there is some suggestions that the jet may have a velocity
structure (but still monodirectional), with a fast spine and a slow layer
(e.g. Laing 1993; Chiaberge et al. 2000 and references therein; 
Ghisellini, Tavecchio \& Chiaberge 2004),
at least in low power (and TeV emitting) BL Lacs and in their
misaligned counterpart, i.e. FR I radiogalaxies.

\subsection{Speeds: GRBs}

We do not see ``jets" in GRBs, but we nevertheless believe that their
emission originates from plasma moving with $\Gamma>100$, making
them the speed record holders among all known sources.
There are essentially three arguments suggesting such large speeds:

\begin{enumerate}
\item
Compactness.  We see variability on timescales
1--100 ms, powers $\sim 10^{50}$ erg s$^{-1}$ and spectra
extending above 511 keV, the threshold for 
$\gamma$--$\gamma \to e^+ e^-$.
If the source is not moving the optical depth for 
the photon--photon process is of the order of $10^{11}$--$10^{12}$.
Relativistic bulk motion comes to the rescue by allowing much
larger sizes of the emitting region than inferred from variability,
much smaller intrinsic powers, and much less photons above threshold.

\item 
Variability. To produce the radiation we see, the source cannot be as small
as inferred from the observed variability, since the required leptons
would make the source Compton or Thomson thick for scattering.
Again, a large $\Gamma$ is required.

\item 
Theory. A huge amount of energy put in a small volume in a short time
generates a small Big--Bang: $e^+$--$e^-$ pairs are immediately created,
the source becomes istantaneously thick and expands, pushed by its
internal pressure, to become relativistic.
This is the {\it fireball} (Cavallo \& Rees 1978).

\end{enumerate}

At first, fireballs were thought to be spherically symmetric, with
radial velocities.
Now there are strong indications that they are collimated
into cones (or flying pancakes) of some aperture angle $\theta_{\rm j}$.
These angles are of the same order of blazar's jets, but in GRBs studies
the ``sprayed' nature of the velocities has been mantained
in almost all models of GRBs (with the exception of the 
cannonball model; e.g. Dar \& De R\`ujula, 2003).

\subsection{Power: blazars}

The spectra (SED) of blazars are characterized by two broad
emission peaks, believed to be produced in the
same zone of the jet (at some hundreds of Schwarzchild radii
from the base) by the synchrotron and inverse Compton 
procesess (but there are alternative views, see e.g. Mucke et al. 2003
and references therein).

\begin{figure}[h]
\begin{tabular}{cc}
\hskip -1 true cm
\psfig{file=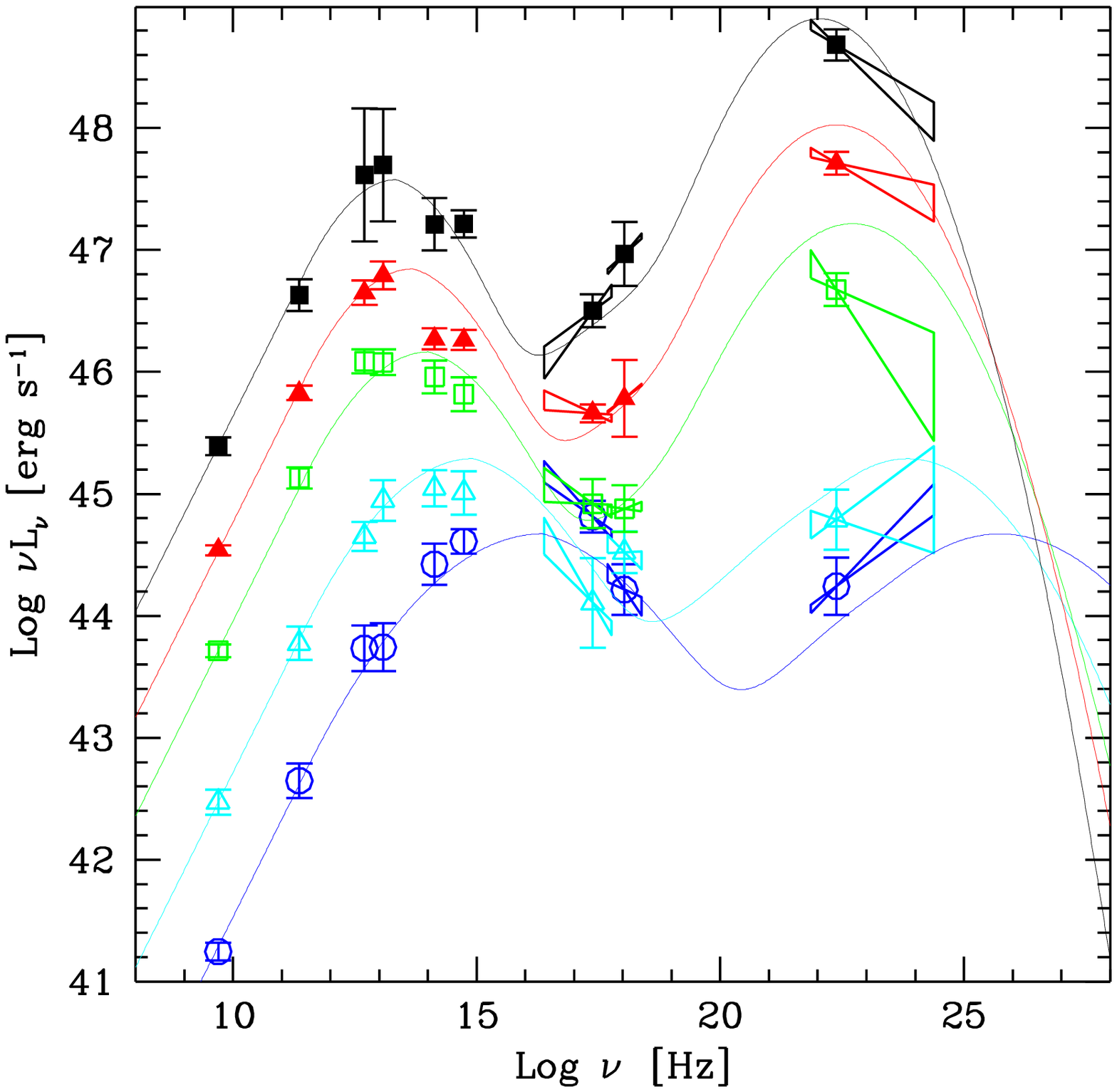, width=7.1cm, height=7.5cm} 
&\hskip -0.7 true cm\psfig{file=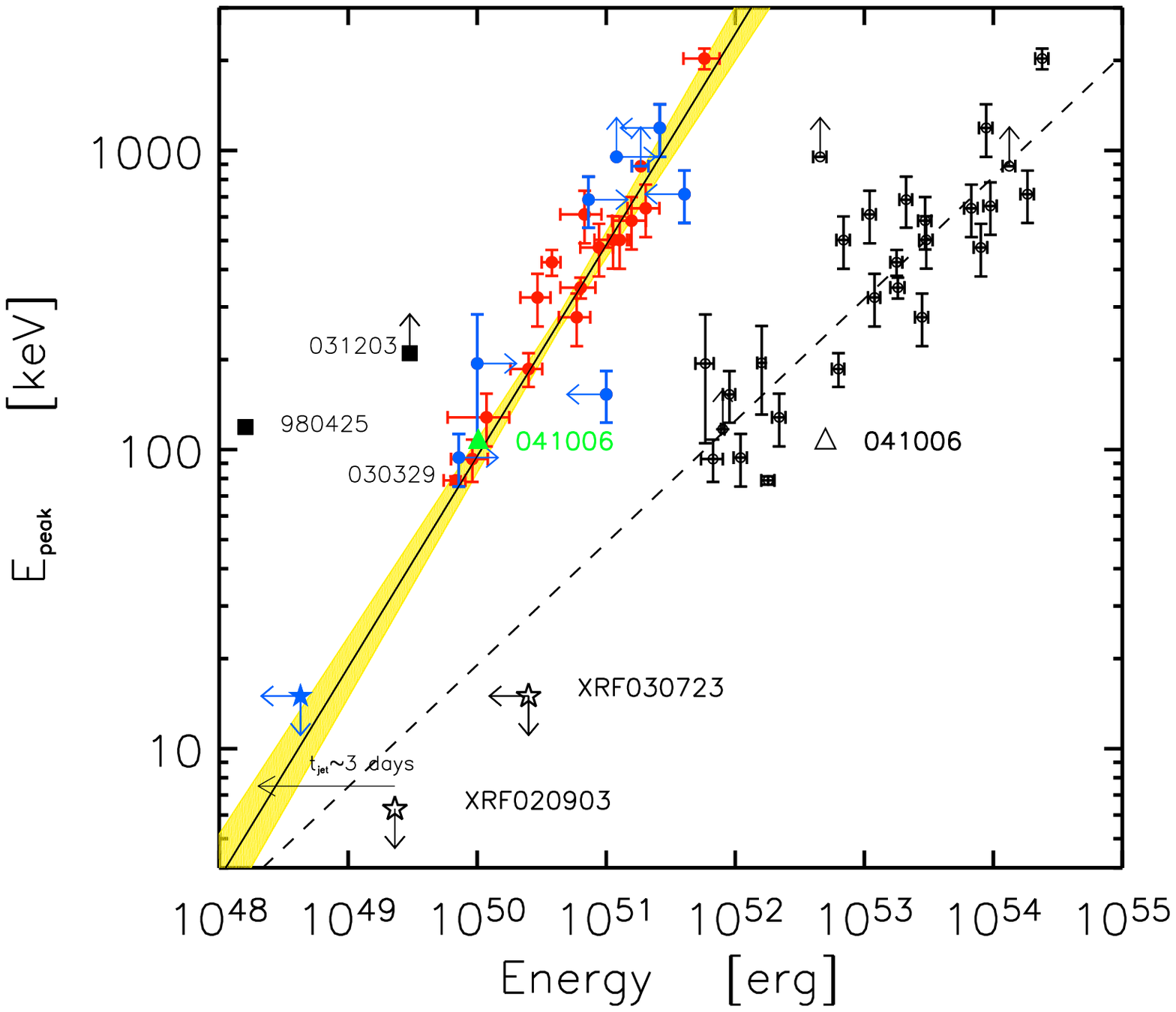, width=7.1 cm, height=7.3cm} 
\end{tabular}
\caption{{\bf Left:} the blazar sequence. From Fossati et al 1998,
Donato et al. 2002. 
{\bf Right:} The GRB sequence (i.e. the ``Ghirlanda correlation"). From
Ghirlanda et al. 2004b, with the addition of the recent GRB 041006;
which nicely fits the correlation. The filled black squares correspond to
the energy emitted during the prompt phase of the bursts 
$E_{\rm iso,\gamma}$ {\it assuming isotropy} (Amati et al. 2002). 
The circles (red and blue) are the {\it collimation corrected} energy 
$E_\gamma=E_{\rm iso,\gamma}(1-\cos\theta)$, where $\theta$ is the aperture
angle of the jet (assumed to be conical).
For both blazars and GRBs the SED is a function of the observed bolometric energy
output, but the two behaviors are opposite:
blazars are bluer when dimmer, GRBs are bluer when brighter.
Note the extremely small scatter of the Ghirlanda correlation.
}
\label{sequence}
\end{figure}

Fossati et al. (1998) found that the SED is controlled by the bolometric
observed luminosity, with both peaks shifting at smaller frequencies when 
the luminosity increases (see Fig. \ref{sequence}, left).
Furthermore, the dominance of the high energy peak increases when
increasing the bolometric luminosity
(but this latter inference was based on the few 
low power BL Lacs detected by EGRET).
This {\it blazar sequence} can be explained by a different degree of
radiative cooling: in powerful blazars electrons cool faster,
producing a break in the electron distribution function
at smaller and smaller energies when increasing the total
(radiation plus magnetic) energy density in the comoving frame
(Ghisellini et al. 1998).

In the most powerful blazars, we believe that most
of the energy carried by the jet is not radiated
away, but it is kept to power the large scale radio--lobes.
Equipartition and minimum energy arguments are used
to infer the energy contained in these structures, 
and by dividing it with the estimated lifetime of the source
the average power of the jet can be estimated.
Rawling \& Saunders (1991) 
find average powers ranging from $10^{43}$--$10^{44}$ erg s$^{-1}$ 
for FR I radiogalaxies to $10^{46}$--$10^{47}$ erg s$^{-1}$ for FR II
radiogalaxies and radio--loud quasars.

One can also calculate the power carried by the jet by
inferring its density through modeling the observed SED
and requiring that the jet carries at least the particles
and the magnetic field necessary to make the radiation we see.
This has been done on the pc scale by Celotti \& Fabian (1993),
on sub--pc scale (the $\gamma$--ray emitting zone) 
by Celotti \& Ghisellini (2004, see also Ghisellini 2003), 
and on the hundreds of kpc scale (the X--ray jets
seen by Chandra) by Celotti, Ghisellini \& Chiaberge (2001)
and Tavecchio et al. (2000).
These studies find consistent values of the power transported
by the jet and require the presence of a dynamically dominating 
proton component (see also the arguments by Sikora \& Madejski 2000).

\subsection{Power: GRBs}

Assuming isotropic emission, GRBs emit 
$E_{\rm \gamma, iso} =10^{51}$--$10^{54}$ erg
in the $\gamma$--ray and hard X--ray band in a few seconds,
and a factor 10 or so {\it less} during the much longer
lasting afterglow.
The prompt spectra are characterized by a single component,
peaking (in $\nu F_\nu$) at an energy $E_{\rm peak}$.
Early results from BATSE led to believe that $E_{\rm peak}$
was clustered around $\sim 300$ keV, but now (mainly thanks
to HETE II, see e.g. Lamb et al. 2004, Sakamoto et al. 2004) 
we think that the $E_{\rm peak}$ distribution
is broader, ranging from a few keV and a few MeV.
Both the prompt and the early
afterglow emissions are very likely collimated
into cones of semiaperture angles $\theta_{\rm j}$,
decreasing the energy budget by the factor $(1-\cos\theta_{\rm j})$
(two jets are assumed; e.g. Frail et al. 2001).
The cone angle can be measured by measuring the time $t_{\rm j}$
at which the afterglow lightcurve steepens.
Being the result of a geometric effect, this break must
be {\it achromatic}.
Recently, Ghirlanda, Ghisellini \& Lazzati (2004) found
a remarkably tight correlation between the 
collimation corrected energy 
$E_\gamma =(1-\cos\theta_{\rm j}) E_{\gamma, {\rm iso}}$ 
and $E_{\rm peak}$ (Fig. \ref{sequence}, right):
$E_{\rm peak}\propto E^{0.7}_\gamma$: then GRBs are  
{\it bluer when brighter}, just the opposite of blazars.
The fact that the scatter around this correlation is
so small has a profound cosmological implication, since
it allows to use GRBs as standard candles (see below).

\section{The central engine}

Despite the efforts of the last 30 years, the acceleration,
launching and collimation of jets are still open issues.
Even the primary energy reservoir is debated,
since it may be in the spin of the black hole,
or in the rotational energy of the accretion disk.
For GRBs the early suggestions of neutrino--neutrino
interactions resulting from the merging of two neutron
stars seems now disfavored by energetic arguments and
by the fact that GBRs preferentially occur well within galaxies.
For blazars, the large ratio between the $\gamma$--ray and 
the X--ray fluxes suggests that the reprocessing due to
$\gamma$--$\gamma \to e^{\pm}$ is not important,
requiring that the jet dissipation region is not
very compact. This in turn requires that most of the 
dissipation takes place at some
distance from the black hole (hundreds of Schwarzchild radii,
Ghisellini \& Madau 1996), 
and energy has to be transported there in a low entropy fashion, 
favoring a Poynting flux origin  of the jet power.
On the other hand, the dominance of the $\gamma$--ray 
emission with respect to the synchrotron radiation 
emitted at lower frequencies requires a magnetic
field which is not dominant, at least in the emission region.
Therefore pure Poynting flux jet models face some difficulties, 
even if the role of the magnetic field at the start of the jet 
is probably crucial.
For GRBs, the ``standard" model 
(e.g. Rees \& M\'esz\'aros 1992; 
Rees \& M\'esz\'aros 1994; 
Sari \& Piran 1997) assumes 
that dissipation takes places right at the start,
with a little ``contamination" of barions,
and that it is the internal pressure that
accelerates the fireball to ultrarelativistic speeds.
However, also for GRBs a low entropy pure Poynting flux has been
proposed (Blandford 2003, Lyutikov \& Blandford 2003).
A possible observational diagnostics for this alternative
model is a high level of polarization 
of the prompt and the early (minutes from the trigger) afterglow 
(Lyutikov, Pariev \& Blandford 2003) which could flag the presence 
of a well ordered and dominant magnetic field.

The most accepted scenario for the production of the prompt
radiation in GRBs proposes that the energy, initially in
high entropy form, is transformed into bulk motion.
To produce radiation, this ordered kinetic energy has to be
re--converted in random energy and then into radiation.
In this scenario this job is done by 
inhomogeneities and different velocities in the ultrarelativistic 
fireball wind which cause {\it internal} shocks.
Generally, these shocks can convert only a modest fraction
of the total energy in radiation, unless the relative bulk
Lorentz factors  of the different parts of the fireball are 
extremely large (Beloborodov 2000, but note that  
this may cause other difficulties, see e.g. Ghisellini 2003).
On the contrary, the interaction of the fireball with the
circumburst medium, which is thought to produce
the afterglow radiation, should be more efficient, since 
the circumburst medium is at rest, and a larger fraction of the
initial kinetic energy of the fireball can be dissipated.
As a result, we naively expect that the total energy
radiated during the afterglow phase is greater than the one
radiated during the prompt phase, contrary to what is observed.

The low efficiency of internal shocks, which is a problem for GRBs, 
is instead required for blazars, where the very same idea of internal shocks
has been successfully applied (Ghisellini 1999; Spada et al. 2001;
Guetta et al. 2004).
Powerful blazars, in fact, radiate a minor fraction of their
jet power, and keep the rest up to the end of the jet, where
it is used to energize the radio lobes.
In addition, internal shocks in blazars explain
why most of the dissipation takes place at some distance
from the black hole: it takes time for a faster part of the
flow to catch up with a slower one: if the two parts
have bulk Lorentz factors $\Gamma$ and $2\Gamma$ and
are initially separated by a distance $R_0$, the collision
will take place at $R_{\rm diss} \sim \Gamma^2 R_0$:
with $R_0$ of the order of the Schwarzchild radius and
$\Gamma\sim 10$, $R_{\rm diss}$ is just right.
After the first collisions,
the range of bulk Lorentz factors is reduced,
and therefore subsequent collisions will take place
with a reduced efficiency: in other words, internal
shocks select a preferred distance where most of the energy
is dissipated, in agreement with what we observe.

\begin{figure}[th]
\centerline{\psfig{file=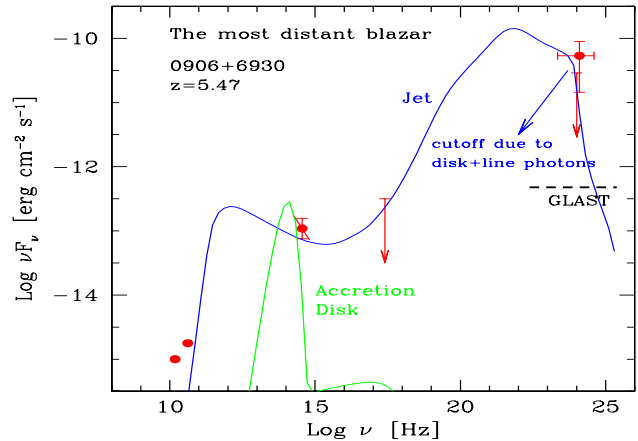, width=9.5cm, height=7cm}}
\vskip -0.5 true cm
\caption{
The Spectral Energy Distribution of the most
distant blazar Q0906+6930 (Romani et al. 2004).
For this and alike blazars, a high energy cutoff is expected in the spectrum
due to the $\gamma$--$\gamma\to e^+ e^-$ process between $\gamma$--rays
and optical--UV photons, both produced in the vicinity of the source
and belonging  to the optical--UV cosmic background photons.
The solid line is a synchrotron inverse Compton model which takes
into account $\gamma$--$\gamma$ absorption, pair creation and
the associated pair reprocessing (along the lines discussed
in Ghisellini et al. 1998).
}
\label{0906}
\end{figure}

\section{Distant blazars and the IR--UV background}

Romani et al. (2004) have recently associated a previously
unidentified EGRET source to the more distant blazar known,
0906+6930, at $z=5.47$. Its SED is shown in Fig. \ref{0906},
together with a possible spectral model.
If confirmed, this would be the most distant $\gamma$--ray emitter.
Besides flagging the presence of really supermassive
(although young) black holes (of masses exceeding $10^{10}M_\odot$),
these sources will be crucial for studying the cosmic
background radiation at UV--Optical frequencies,
which will imprint a sharp cutoff at high (tens of GeV)
$\gamma$--ray energies.
Similarly, but at much lower redshifts, one can study the IR
background through the absorption it produces  at TeV energies.
One should be careful to disantangle the {\it local} absorption
due to photons produced within the source (which, incidentally,
must also act as seed photons for the scattering process) from
the absorption due to the cosmic backgrounds, 
but this is possible by studying sources of similar luminosities
at different redshifts. 
In this respect, one can judge the potential of GLAST by looking
its detection limit in Fig. \ref{0906}.

\section{Measuring the Universe with GRBs}
\begin{figure}[th]
\centerline{\psfig{file=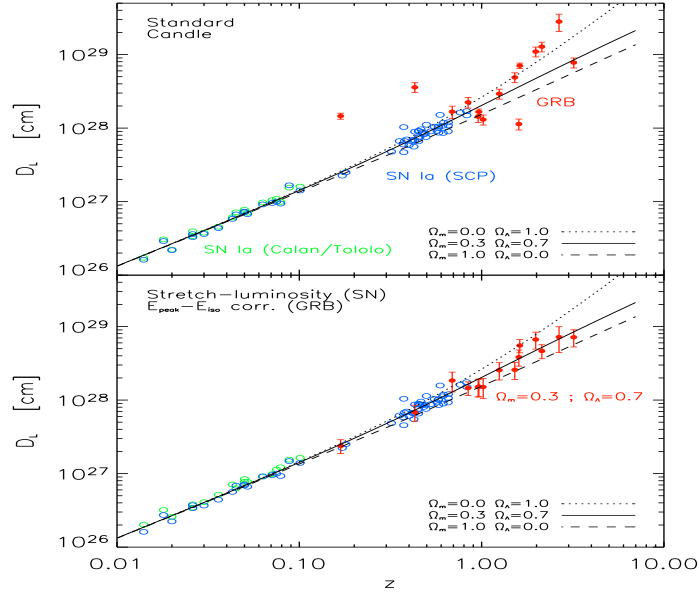, width=9cm, height=8cm}}
\caption{Classical Hubble  diagram in the
form of  luminosity--distance $D_{\rm L}$ vs redshift $z$ for Supernova Ia
and GRBs (filled  red circles: the 15 bursts in Ghirlanda et al 2004b).  
In the top panel no  correction was applied for the SN Ia  and  all GRBs 
are assumed to emit $E_\gamma=10^{51}$ erg.  
In the bottom  panel  we  have  applied the  stretching-luminosity  and  the
$E_\gamma$--$E_{\rm peak}$ relations  to SN Ia and  GRBs, respectively.  
Both panels also show curves for different $D_{\rm L}(z)$, as labelled. From
Ghirlanda et al. (2004a).
 }
\label{hd}
\end{figure}
\begin{figure}[h]
\begin{tabular}{cc}
\hskip -1.2 true cm
\psfig{file=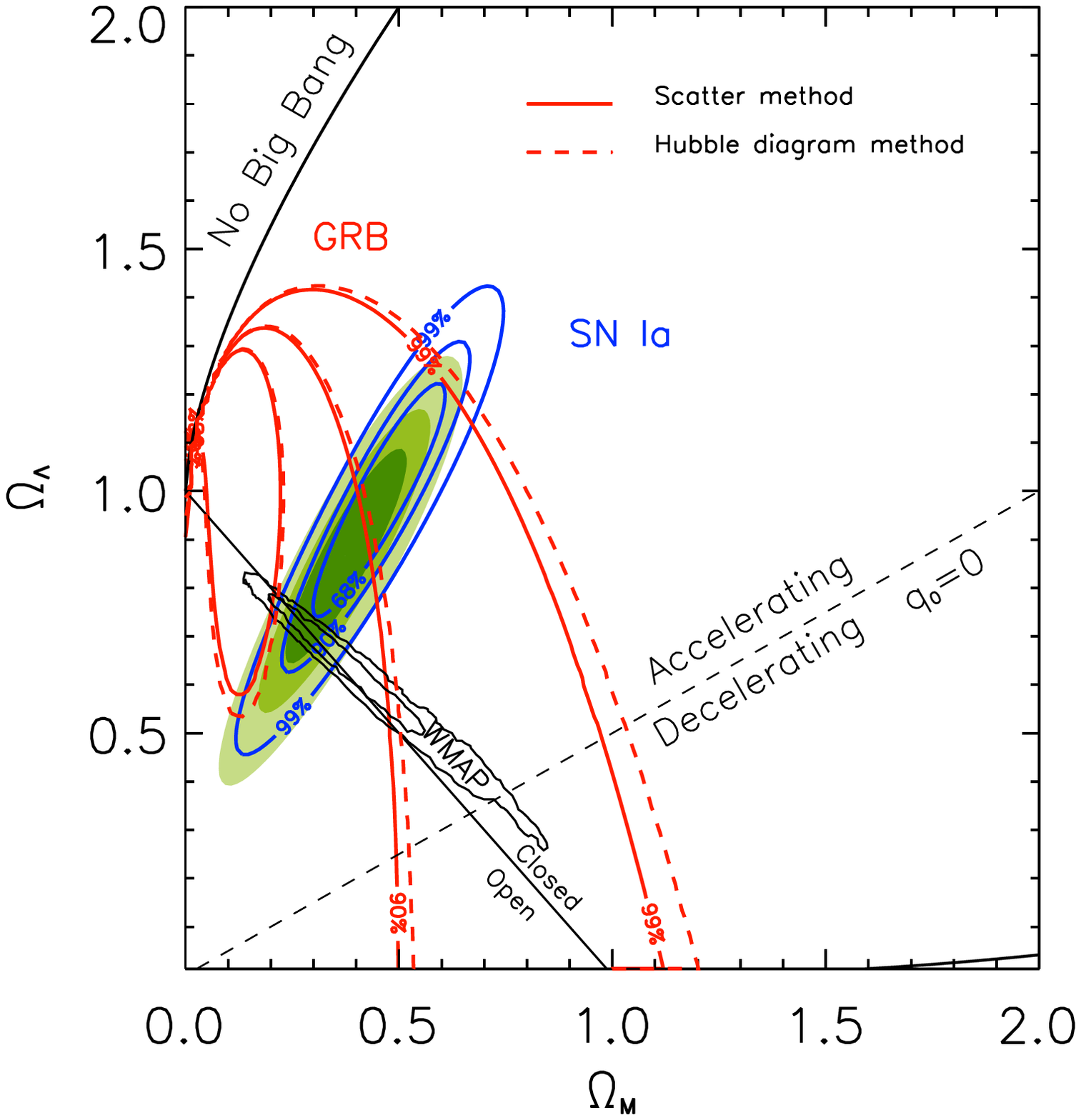, width=7.3cm, height=7cm} 
&\hskip -0.9 true cm\psfig{file=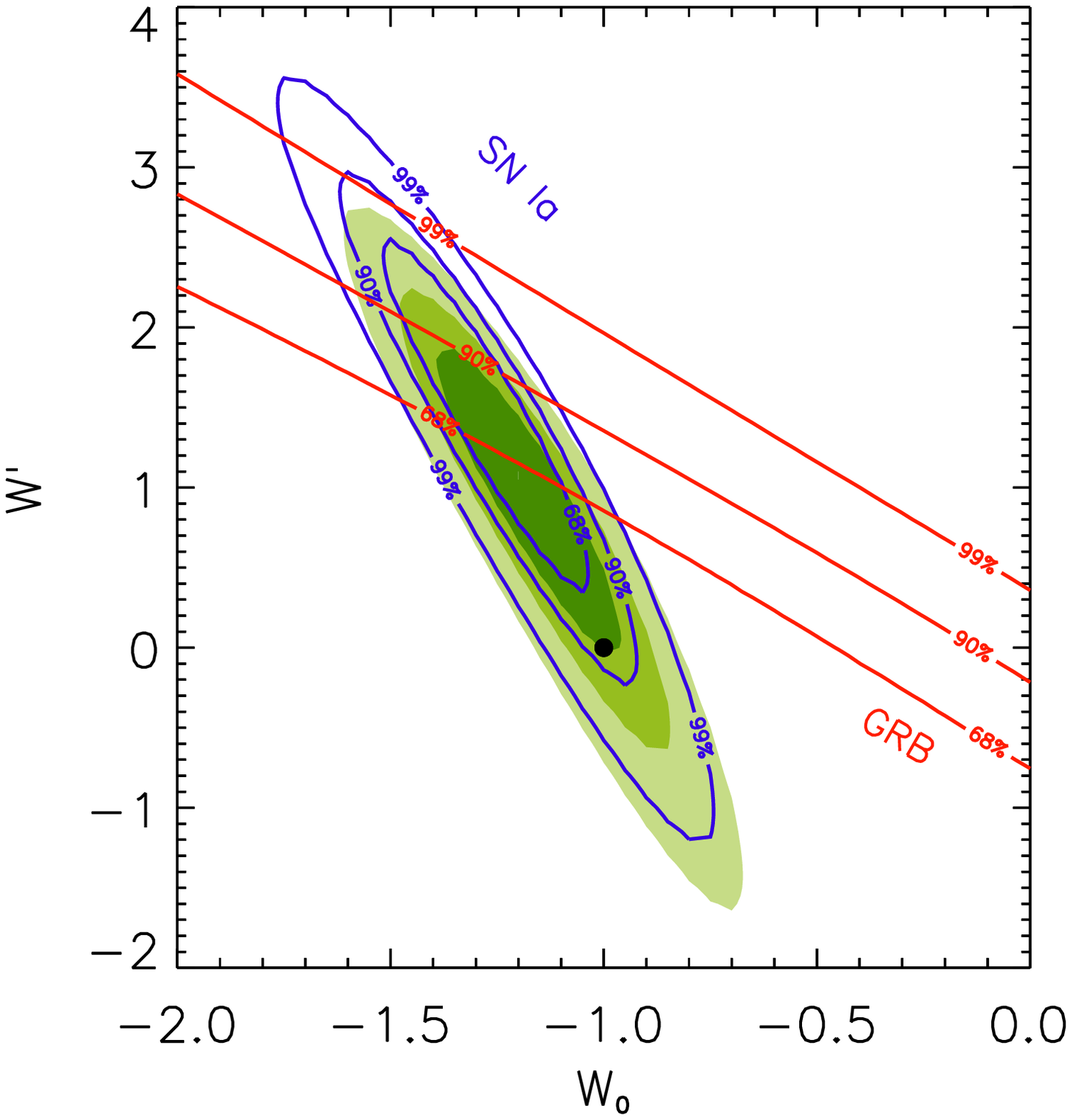, width=7.3cm, height=7cm} 
\end{tabular}
\caption{{\bf Left:} Constraints in the 
$\Omega_{\rm M}$--$\Omega_{\Lambda}$ plane derived for 
the GRB sample (15 objects, red contours); 
the ``Gold" Supernova Ia sample of Rieass et al. (2004)
(156 objects, blue contours). 
The WMAP satellite constraints (black contours, Spergel et al. 2003) 
are also shown.
The three colored ellipsoids are the confidence regions (dark green: 68\%; 
green: 90\%; light green: 99\%) for the combined fit of SN Ia 
and our GRB sample. 
{\bf Right:} Constraints in the $w^\prime$--$w_0$ plane,
where the equation of state of the dark energy is
assumed to be described by $P=(w_0+w^\prime z)\rho c^2$.
The combined GRB+SN Ia contours are more consistent
with $w_0=-1$, $w^\prime=0$ than SN Ia alone. From Ghirlanda et al. 2004a.
}
\label{cosmo}
\end{figure}

The ``Ghirlanda correlation" shown in the right 
panel of Fig. \ref{sequence} opens up the possibility
to use GRBs as rulers to measure the universe.
This is more than a wishful thinking, and already
now the 15 GRBs of the sample of Ghirlanda et al. (2004b)
put interesting constraints on the cosmological parameters.
Fig. \ref{hd} shows the classical Hubble diagram
for both SN Ia and GRBs.
In the top panel we can see the data points before correcting
the luminosity of SN Ia for the Phyllips relation (i.e. a relation
connecting the rate of decay of the lightcurve with the luminosity
of the event; Phyllips 1993), and assuming that all GRBs emit $10^{51}$ erg.
In the bottom panel, both the luminosity of SN Ia and the energetics
of GRBs have been corrected, assuming, for GRBs, the
Ghirlanda relation between $E_\gamma$ and $E_{\rm peak}$.
The left panel of Fig. \ref{cosmo} shows the constraints in the 
$\Omega_{\rm M}$--$\Omega_{\Lambda}$ plane, considering
GRBs alone, SN Ia alone, and a combined fit.
One can see that despite the huge difference in number (15 GRBs
vs 156 SN Ia), GRBs can affect the constrained derived
by SN Ia, to make the combined fit more in agreement with
the ``concordance" cosmological model, which suggests
$\Omega_{\rm M}\sim 0.3$ and $\Omega_{\Lambda}\sim 0.7$.
But the huge potential of GRBs for cosmology is not much
in finding more accurate values of $\Omega_{\rm M}$ and $\Omega_{\Lambda}$;
i.e. in saying how the universe is now.
The real potential lies in the ability of GRBs 
to find out in which way the universe became what it is,
i.e. how the dark energy evolves.
To do that we must have sources of known luminosity or power
embracing the maximum possible range of redshifts, to find
out, for instance, if the equation of state of the
dark energy was described by a constant or a time dependent
relation between pressure and energy density.

For illustration, the right panel of Fig. \ref{cosmo} shows the constraints
in the $w^\prime$--$w_0$ plane, when  assuming a time dependent
equation of state of the dark energy, of the form
$P = (w_0+w^\prime z) \rho c^2$, where $P$ and $\rho$ are the pressure
and the energy density of the dark energy. 
If the dark energy is described by the classical $\Lambda$ cosmological
term, then we must have $w_0=-1$ and $w^\prime=0$.
Again, note that although GRBs alone are not very constraining,
the combined fit of GRB+SN Ia moves the 68\% confidence region
to embrace the $w_0=-1$, $w^\prime=0$ point.

An obvious advantage from using GRBs for cosmology is that 
we can detect them up to redhisft 10 or 20, contrary 
to SN Ia, detectable up to $z\sim 1.7$.
The other point in favor is that GRBs are virtually unaffected
by dust or Lyman $\alpha$ absorption.
Finally, if the Ghirlanda correlation is obeyed by GRBs
of any luminosity, it is evolution--independent.

The SWIFT satellite is expected to localize 100--150 bursts per year.
For the majority of them  we foresee to have redshifts
and well sampled afterglow lightcurves, allowing
the most optimistic expectations for doing cosmology
with GRBs.


\section*{Acknowledgements}
It is a pleasure to thank A. Celotti,
C. Firmani,
G. Ghirlanda, 
D. Lazzati and
F. Tavecchio 
for many helpful discussions and a 
fruitful collaboration.


\end{document}